\documentclass[preprint,showkeys,aps,prb]{revtex4}
\usepackage[dvips]{graphicx}

\begin{document}

\title{Aspects of Nos\'e and Nos\'e-Hoover Dynamics Elucidated}

\author{William Graham Hoover and Carol Griswold Hoover,\\ Ruby Valley Research Institute,
601 Highway Contract 60,\\ Ruby Valley, Nevada 89833}

\date{\today}

\keywords{Nos\'e Oscillator, Nos\'e-Hoover Oscillator, Dettmann Oscillator, Nonlinear Dynamics}
\vspace{0.1cm}

\begin{abstract}
Some paradoxical aspects of the Nos\'e and Nos\'e-Hoover dynamics of 1984 and Dettmann's 
dynamics of 1996 are elucidated. Phase-space descriptions of thermostated harmonic oscillator
dynamics can be simultaneously expanding, incompressible, or contracting, as is described
here by a variety of three- and four-dimensional phase-space models. These findings illustrate
some surprising consequences when Liouville's continuity equation is applied to Hamiltonian flows.
\end{abstract}

\maketitle

\section{Introduction to Harmonic Oscillator Models}

In 1984 Shuichi Nos\'e noticed that introducing a scale factor $s$ into the momenta made it
possible to convert Gibbs' microcanonical distribution to the canonical one\cite{b1,b2} provided
that [ 1 ] the ``time-scaling factor'' $s$ was governed by a temperature-dependent logarithmic potential
and [ 2 ] the equations of motion ( assumed ergodic ) for the coordinates $\{ \ q \ \}$ and momenta
$\{ \ p \ \}$ were multiplied by $s$. In principle and in practice this development made it possible
to generate a dynamics consistent with the canonical ensemble, for systems small and large. And for
ergodic systems such a dynamics, with the weights of dynamical states given by the Boltzmann factor
$f(q,p) \propto e^{-{\cal H}/kT}$, can closely approximate the predictions of Gibbs' canonical ensemble.

Hoover soon pointed out that the time-scaling factor $s$ was completely extraneous.  He showed that the
very same isothermal equations of motion could be derived directly from the phase space continuity equation,
$$
(\dot f/f) = -(\dot \otimes/\otimes) \equiv -\sum [ \ (\partial \dot q/\partial q) + (\partial \dot p/\partial p) \ ]
$$
without introducing time scaling\cite{b3,b4}. Here $f$ is probability density and $\otimes$ represents an
infinitesimal comoving phase volume. In Hoover's adaptation of Nos\'e's approach,``Nos\'e-Hoover dynamics'',
the momentum conjugate to $s$ appears as a friction coefficient $\zeta$. $\zeta$ controls the dynamics
{\it via} integral feedback using a target value of the kinetic temperature $mkT \equiv \langle \ p^2 \
\rangle$. Here $k$ is Boltzmann's constant and $m$ is a particle's mass.  Hoover found that the oscillator was
far from ergodic.  With Posch and Vesely he demonstrated the presence of a modest chaotic sea ( six percent of
the stationary solution ) for the oscillator in addition to the preponderant ( 94\% ) quasiperiodic toroidal
solutions\cite{b4,b5}.  For the oscillator Nos\'e's Hamiltonian ( with $\zeta = p_s$ ) is :
$$
{\cal H}_{Nos\acute{e}} = [ \ q^2 + (p/s)^2 + \ln(s^2) + \zeta^2 \ ]/2 \ .
$$
For simplicity we choose $m$, $k$, and the oscillator force constant all equal to unity.

In 1996 Carl Dettmann showed that the Nos\'e-Hoover equations of motion can be derived from a
{\it scaled} Hamiltonian, provided that the energy itself is set equal to zero\cite{b6,b7}.
$$
{\cal H}_{Dettmann} \equiv s{\cal H}_{Nos\acute{e}} \equiv 0 \ .
$$
These novel and surprising ideas are most easily displayed, illustrated, and understood by applying
them again to the simplest possible problem, a one-dimensional harmonic oscillator\cite{b3,b4}.
Despite its lack of ergodicity such an oscillator provides a surprising source of topological variety,
including intricately knotted phase-space trajectories ! Thus it has captured the attention of
mathematicians as well as physicists and chemists\cite{b8,b9,b10,b11,b12,b13}.

Our purpose here is entirely pedagogical.  We focus on some surprising qualitative differences among the
three- and four-dimensional flows described by Nos\'e, Dettmann, and Nos\'e-Hoover dynamics. Each of them
can be analyzed in a four-dimensional $(q,p,s,\zeta)$ phase space, or in a three-dimensional subspace
corresponding to the restriction of constant energy. Liouville's Theorem, that Hamiltonian flows are
incompressible, is a straightforward consequence of the motion equations :
$$
\{ \ \dot q = (\partial {\cal H}/\partial p) \ ; \ \dot p + -(\partial {\cal H}/\partial q) \ \} \longrightarrow
\dot f = (\partial f/\partial t) + \sum \dot q(\partial f/\partial q) + \dot p(\partial f/\partial p) \equiv 0 \ .
$$
The Nos\'e ($s^0$) and Dettmann ($s^1$) oscillator Hamiltonians differ by just a factor $s$ :
$$
{\cal H}_{N,D} = (s^{0,1}/2)[ \ q^2 + (p/s)^2 + \ln(s^2) + \zeta^2 \ ] \equiv 0 \ ;
 \ \zeta \equiv p_s \ .
$$
In both cases the resulting constant-energy dynamics develop in a three-dimensional constrained phase
space. For instance we can choose a space described by the coordinate $q$, scaled momentum $(p/s)$, and
friction coefficient $\zeta$.  With the energy fixed any one of the four variables $(q,p,s,\zeta)$ can 
be determined from a convenient form of the constraint conditions :
$$
s = e^{ -(1/2)[ \ q^2+(p/s)^2+\zeta^2 \ ] } \ .
$$
It is convenient to specify $(q,p/s,\zeta)$ and then to select $s$ to satisfy the ${\cal H} \equiv 0$
constraints. A consequence of the Dettmann multiplier $s^1$ is the simple relationship linking solutions
of the Nos\'e and Dettmann Hamiltonians :
$$
(\dot q,\textstyle{\frac{d}{dt}}(p/s),\dot \zeta)_{Dettmann} \equiv s(\dot q,\textstyle{\frac{d}{dt}}(p/s),\dot \zeta)_{Nos\acute{e}} \ .
$$
The Nos\'e and Dettmann trajectories are identical in shape but are traveled at different speeds.

\begin{figure}
\includegraphics[width=2.5in,angle=90.]{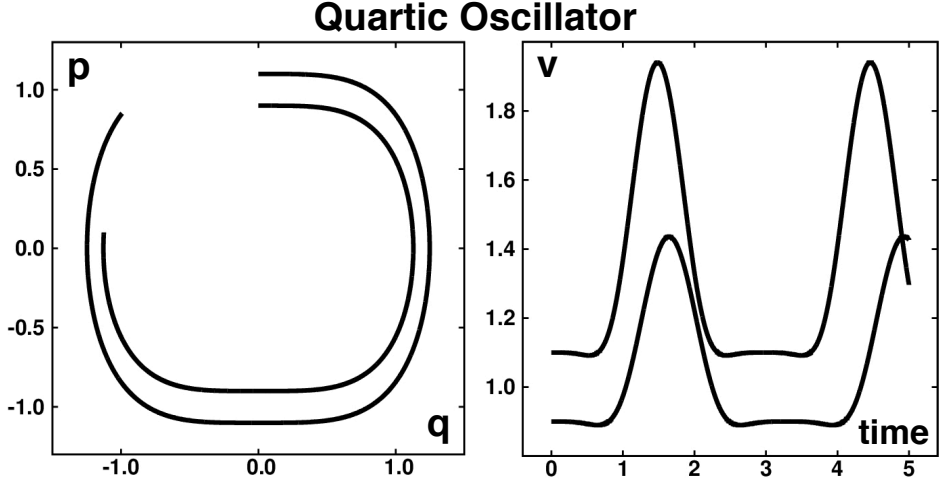}
\caption{
To the left we see two quartic oscillator trajectories in the $(q,p)$ phase space. The initial states
are $(0,0.9)$ and $(0,1.1)$. The trajectories include 5000 fourth-order Runge-Kutta timesteps with
$dt = 0.001$. The Hamiltonian is ${\cal H} = (q^4/4)+(p^2/2)$. Because frequency increases with energy
a constant comoving area shears as time progresses.  Although both $f(q,p,t)$ and $\otimes(q,p,t)$ are
constants of the motion, obeying Liouville's Theorem, the phase-space speed, $\sqrt{\dot q^2 + \dot p^2}
= \sqrt{p^2 + q^6}$ is far from constant, as is shown to the right in the Figure for the two trajectories..
}
\end{figure}

It is tempting to think that the time spent in the volume element $dqd(p/s)d\zeta$ in the Dettmann case is 
proportional to $(1/s) \equiv e^{+(q^2 + (p/s)^2 + \zeta^2)/2}$ compared to the Nos\'e case.  But this
relative probability of $e^{+{\cal H}/kT}$ is the {\it reciprocal} of what we would ( naively ) expect.
Evidently the time argument is false. To see why, consider the Hamiltonian motion of a {\it quartic} oscillator
with ${\cal H} = (p^/2) + (q^4/4)$ .  Both the phase-space trajectory and the phase-space speed,
$\sqrt{\dot q^2 + \dot p^2}$ are shown in {\bf Figure 1}. Though the Hamiltonian is constant, the speed in
phase space, $\sqrt{ q^6 + p^2 }$ varies. Liouville's Theorem correctly shows that the probability density $f$
and the comoving area $\otimes$ are both constants along a trajectory.  But, because the {\it shape} of the area
varies with time there is no simple link between speed and $f$ or $\otimes$.  It is the changing width of
a comoving element perpendicular to the trajectory that destroys the supposed  connection between speed
and probability.

Our goal here is simply to point out this complex relationship between speed and probability in the simplest
possible example. The difference can be even more dramatic in three- and four-dimensional problems. Let us
look at the simplest such example problem in order to enrich our understanding.  Consider the smallest periodic
orbit traced out by the Dettmann, Nos\'e, and Nos\'e-Hoover equations of motion.  We choose to begin the orbit
with a higher kinetic energy, $1.55^2/2$, than the target value of $1/2$.  With the initial conditions
$(q,p/s,\zeta) = (0,1.55,0)$ we find $s = \sqrt{e^{-1.55^2}} = 0.30082 \rightarrow p = 0.46627$ so that
the initial condition $(q,p,s,\zeta) = (0,0.46627,0.30082,0) \rightarrow {\cal H} = 0$. 

\section{An expanding model in four dimensions}

Nos\'e's Hamiltonian, ${\cal H}_N = (1/2)[ \ q^2 + (p/s)^2 + \ln(s^2) + \zeta^2 \ ]$, followed by
time-scaling, leads to four equations of motion in $(q,p,s,\zeta)$ space:
$$
\{ \ \dot q = p/s \ ; \ \dot p = -sq \ ; \ \dot s = s\zeta \ ; \
\dot \zeta = [ \ (p/s)^2 - 1 \ ] \ \} \ \rightarrow (\partial \dot s/\partial s) = +\zeta \ .
$$
Exactly these same motion equations follow more simply from Dettmann's Hamiltonian, with no need of
time scaling. Because our initial condition has a higher ``temperature'' $\langle \ (p/s)^2 \ \rangle$
than the target of unity, the short-time friction coefficient $\zeta$ becomes positive, suggesting,
from $\dot s = s\zeta$ that Nos\'e's (or Dettmann's ) oscillator's phase volume begins by expanding
rather than contracting. This expansion with a positive friction seems counter to Liouville's Theorem,
suggesting a paradox. {\bf Figure 2} shows the details of this four-dimensional problem. The time
scaling factor $s$ {\it is} precisely equal to Gibbs' canonical probability density. With the
short-time positive friction, $\zeta > 0$, the flow does contract rather than expand. Let us investigate
this intriguing problem further.

\begin{figure}
\includegraphics[width=2in,angle=+90.]{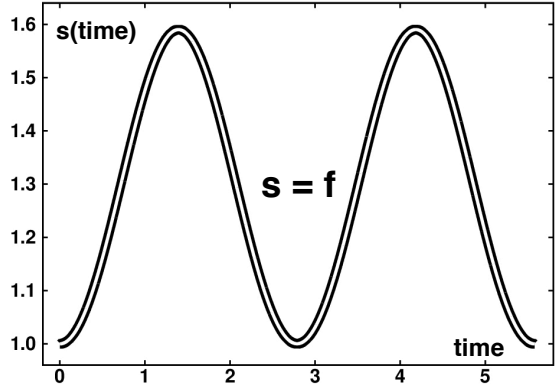}
\caption{
The time variation of two expressions for the probability density $f$ as measured once around a periodic
orbit generated with Dettmann's (or Nos\'e's, with time scaling) Hamiltonian in the four-dimensional
$(q,p,s,\zeta)$ phase space.  The initial conditions are (0, 0.46627, 0.30082, 0) so that initially the
scaled momentum is $(p/s) = 1.55$ and the Hamiltonian vanishes. The thicker line is Gibbs' canonical-ensemble
density $e^{-[ \ q^2 + (p/s)^2 + \zeta^2 - 1.55^2 \ ]/2}$. The thinner white line overlaying the thicker black
one shows the progress of the ``time-scaling factor'' $s(t)/s(0) = e^{\int_0^t\zeta(t')dt'}$. The perfect
agreement demonstrates that the phase-space density $f(q,p,\zeta)$ {\it can} be obtained by measuring the
phase-space compression ( but not the speed ) along the four-dimensional Hamiltonian trajectory with
Dettmann's constraint, ${\cal H}_D \equiv 0$ .  But the early-time association of increasing phase volume,
expected from $(\partial \dot s/\partial s) = \zeta > 0$, is indeed paradoxical.
}
\end{figure}

\section{An incompressible model ?}

Dettmann's Hamiltonian, ${\cal H}_D = (s/2)[ \ q^2 + (p/s)^2 + \ln(s^2) + \zeta^2 \ ]$, with the constraint
${\cal H}_D\equiv 0$ imposed in the initial conditions, is not really {\it incompressible} :
$$
\{ \ \dot q = p/s \ ; \ \dot p = -sq \ ; \ \dot s = s\zeta \ ; \
\dot \zeta = -(1/2)[ \ q^2 - (p/s)^2 + \ln(s^2) + \zeta^2 \ ] - 1 \ \} \ \rightarrow
$$
$$
(\partial \dot s/\partial s) + (\partial \dot \zeta/\partial \zeta) = + \zeta - \zeta = 0 \ 
[ \ {\rm Incompressible?} \ ] \ .
$$

The flow equations certainly maintain a comoving {\it four}-dimensional hypervolume {\it unchanged in size}. This
is nothing more  than the usual application of Liouville's Theorem and is no surprise.  But taking the zero
energy constraint into account reduces the flow to three phase-space dimensions, as in the Nos\'e-Hoover
picture.  Let us look at that picture next.  The quantitative details of the evolving phase probability are
shown in {\bf Figure 3}.

\section{A contracting model in three dimensions}

Here either Nos\'e-Hoover dynamics or a three-dimensional version of Dettmann's Hamiltonian,
including the constant-energy constraint, gives the same results. A time-reversible frictional
force, $-\zeta p$, provides a steady-state Gaussian phase-space distribution
$e^{-[ \ q^2+p^2+\zeta^2 \ ]/2}$ .
In the two versions of dynamics the friction coefficient $\zeta$ is determined by integral feedback :
$$
\{ \ \dot q = p \ ; \ \dot p = -q - \zeta p \ ; \ \dot \zeta = p^2 - 1 \ \} \ \longrightarrow
(\partial \dot p/\partial p) = -\zeta \ .
$$
Dettmann's motion equations are identical to these if his scaled momentum $(p/s)$ is replaced by the
symbol $p$. Here, with the relatively ``hot'' initial condition, the three-dimensional phase-space
volume {\it shrinks} (correctly) initially due to contraction parallel to the momentum axis. So, for
the three phase-space descriptions of the same physical problem we have found expansion,
incompressibility, and compression, all for exactly the same phase-space states.  We put these three
examples forward from the standpoint of pedagogy, as a useful and memorable introduction to the
significance of Liouville's Theorem for isoenergetic flows. The constraint of constant energy can
lead to qualitative differences in the evolution of $f$ and $\otimes$.

\begin{figure}
\includegraphics[width=2in,angle=+90.]{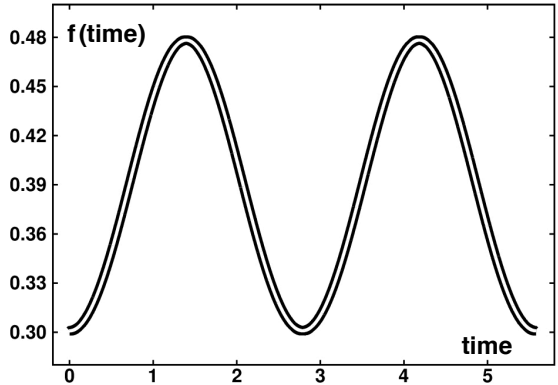}
\caption{
Two probability densities as measured once around a periodic Nos\'e-Hoover orbit in three-dimensional
$(q,p,\zeta)$ space.  The initial values are $(q,p,\zeta) = (0, 1.55, 0)$.  The thicker line is
Gibbs' canonical $e^{-(q^2 + p^2 + \zeta^2)/2}$.  The overlaying thinner white line is
$e^{\int_0^t\zeta(t')dt'}e^{-1.55^2/2}$.  Here the perfect agreement shows that the integrated
three-dimensional phase-space compression corresponds precisely to Gibbs' canonical distribution.
}
\end{figure}

\section{Acknowledgements}
Over 35 years we have studied the paradoxical aspects of Shuichi Nos\'e's pioneering
contribution to statistical mechanics and nonlinear dynamics.  During this time we have benefitted
from the insights of dozens of colleagues throughout the world, most recently from Puneet Patra,
Clint Sprott, Lei Wang, and Xiao-Song Yang, but dating back to Nos\'e himself plus Harald Posch and
Franz Vesely. We thank them all.

\end{document}